\documentclass[onecolumn,prl,superscriptaddress]{revtex4}

\usepackage{graphicx}% Include figure files
\usepackage{dcolumn}% Aligntable columns on decimal point
\usepackage{multirow}
\usepackage{bm}% bold math
\usepackage{times}
\usepackage{color}
\usepackage{amstext}
\usepackage{amsmath}
\usepackage{amsfonts}

\begin{document}

\makeatletter
\long\def\@makecaption#1#2{%
  \par
  \vskip\abovecaptionskip
  \begingroup
   \small\rmfamily
   \sbox\@tempboxa{%
    \let\\\heading@cr
    #1 (color online). #2\hskip1pt%
   }%
   \@ifdim{\wd\@tempboxa >\hsize}{%
    \begingroup
     \samepage
     \flushing
     \let\footnote\@footnotemark@gobble
     #1 (color online). #2\hskip1pt\par
    \endgroup
   }{%
     \global \@minipagefalse
     \hb@xt@\hsize{\hfil\unhbox\@tempboxa\hfil}%
   }%
  \endgroup
  \vskip\belowcaptionskip
}%
\makeatother

\title{Comment on "Infrared signature of the superconducting gap symmetry in iron-arsenide superconductors", Y. M. Dai et al. arXiv:1106.4430}

\author{A. Charnukha}
\email{a.charnukha@fkf.mpg.de}
\affiliation{Max-Planck-Institut f\"ur Festk\"orperforschung, Heisenbergstrasse 1, D-70569 Stuttgart, Germany}
\author{O. V. Dolgov}
\affiliation{Max-Planck-Institut f\"ur Festk\"orperforschung, Heisenbergstrasse 1, D-70569 Stuttgart, Germany}
\author{A. A. Golubov}
\affiliation{Faculty of Science and Technology and MESA+ Institute of Nanotechnology, 7500 AE Enschede, The Netherlands}
\author{B. Keimer}
\affiliation{Max-Planck-Institut f\"ur Festk\"orperforschung, Heisenbergstrasse 1, D-70569 Stuttgart, Germany}
\author{A. V. Boris}
\email{a.boris@fkf.mpg.de}
\affiliation{Max-Planck-Institut f\"ur Festk\"orperforschung, Heisenbergstrasse 1, D-70569 Stuttgart, Germany}

\begin{abstract}
\end{abstract}

\maketitle
Y. M. Dai et al. in their recent work~\cite{Dai_infrared_2011} presented a reflectivity study of the in-plance optical conductivity of a $\textrm{Ba}_{0.6}\textrm{K}_{0.4}\textrm{Fe}_2\textrm{As}_2$ (BKFA) superconductor with $T_{\textrm{c}}\approx39K$. The single crystals used in this study are of high quality and the measurements are rather accurate. The authors analyzed the optical conductivity of BKFA in the framework of the Mattis-Bardeen theory and the BCS theory of superconductivity. This analysis, however, generates a series of severe contradictions with a large body of experimental data obtained with multiple probes as well as misuses both the Mattis-Bardeen and the BCS theories of superconductivity beyond their approximation regions.
\par 1) The authors extracted two superconducting gaps $\Delta_0^{\textrm{S}}=10\ \textrm{meV}$ and $\Delta_0^{\textrm{L}}=16.5\ \textrm{meV}$. It is now clear from numerous ARPES~\cite{PhysRevB.79.054517,PhysRevLett.105.117003,PhysRevB.83.020501,XuDing_ARPES_3D_2011,Shimojima_Science_332.564.2011,Evtushinsky.bogolubons2011}, specific heat~\cite{PhysRevLett.105.027003} and STM measurements~\cite{Shan-Wen_STM_2011} that there is a superconducting gap in BKFA at a much smaller energy, $\Delta_0^{\textrm{S}}=3-4\ \textrm{meV}$. The authors disregard the presence of this smaller gap thus contradicting a whole body of current experimental results. Even assuming that the respective band contributes only 15\% to the total spectral weight of the itinerant response, incorporation of such an additional Mattis-Bardeen-type conductivity term into the authors' analysis would render the description inadequate.
\par 2) The authors introduced a large gap of 16.5 meV, never before observed with any experimental technique including Refs.~6,7,19-23 in their own report. The statement that the values of the superconducting gaps obtained by the authors are well within the reported range for this material is incorrect.
\par 3) Both gap ratios $2\Delta/k_{\textrm{B}}T_{\textrm{c}}=5.9\ \textrm{and}\ 9.8$ obtained by the authors are significantly higher than the BCS weak-coupling limit, which leads the authors to the conclusion of the strong coupling in this compound. At the same time, these gap ratios were obtained within the Mattis-Bardeen theory, which is a {\it weak-coupling} extension of the {\it weak-coupling} BCS theory to finite impurity scattering. Neither of them, in the case of strong interband coupling, provides an adequate description of the superconducting properties even in the low-coupling limit~\cite{PhysRevB.79.060502}. Thus the authors arrived at an obvious contradiction by inferring strong-coupling superconductivity within a weak-coupling theory. An appropriate description of the properties of strongly-coupled superconductors requires taking into account the electron-phonon renormalization, as is implemented in the Eliashberg theory. This makes the formalism more complicated and allows for an analytical solution only in the limit of the Mattis-Bardeen theory, $\gamma\gg\lambda\Omega$, where $\lambda$ is the coupling constant to the intermediate boson with a characteristic frequency $\Omega$. The authors maintain that the strongly-coupled BKFA is in the clean limit because the potassium atoms are doped between the superconducting planes and do not introduce an additional scattering potential. This clearly places BKFA beyond the Mattis-Bardeen approximation. It is also worth noticing that the gap ratios obtained in this report stand out from the general trend observed in Ref.~\onlinecite{Inosov.supergap2011} based on the available experimental data for most of the known conventional and high-temperature superconductors studied during the last three decades.
\par 4) The authors' conclusion of strong-coupling superconductivity is based on the observation that the temperature dependence of the smaller superconducting gap follows that of the larger. The authors appear to be unaware of the fact that within the Eliashberg theory the far-infrared conductivity is suppressed at energies $2\Delta^{\textrm{L}}_0+\Omega$, much larger than the superconducting gap $2\Delta^{\textrm{L}}_0$. The authors explain this behavior by introducing the $16.5\ \textrm{meV}$ superconducting gap. The identical temperature dependence of the optical conductivity at these energies, which is highlighted by the authors as a combined behavior of two separate gaps, is obvious in the Eliashberg formalism.
\par All these issues have already been reconciled in Ref.~\onlinecite{Charnukha.Eliashberg2011}, which reports closely similar data and is inadequately cited by the authors. The theoretical analysis in this work employs a full four-band strong-coupling Eliashberg approach taking into account all of the experimental results available for BKFA. This work also provided a justification for using two-band models with strong interband- and weak intraband-coupling superconductors for the description of the optical conductivity of iron pnictides. It already attributed the qualitative differences in the far-infrared conductivity of BKFA and BFCA in the superconducting state to the different impurity scattering rates, namely, low for BKFA and large for BFCA, based on the fact that cobalt is doped into the superconducting planes but not potassium. The same explanation is reused by the authors without proper reference.
\par It is important to note that the experimental data obtained by Y. M. Dai et al. are virtually identical to those in Ref.~\onlinecite{Charnukha.Eliashberg2011}, as well as in Ref.~\onlinecite{PhysRevLett.101.107004}, cited by the authors. We conclude that the contradiction in the description of the infrared properties of BKFA come from an inadequate theoretical approach adopted by Y. M. Dai et al. and disregard of the existing experimental data.\eject
%------------- THEBIBLIOGRAPHY ------------------------

%------------- THEBIBLIOGRAPHY ------------------------
\end{document}